\documentclass[%
 reprint,
superscriptaddress,
 aps,
 prl,
]{revtex4-2}

\usepackage{graphicx}
\usepackage{dcolumn}
\usepackage{multirow}
\usepackage{booktabs}
\usepackage{multirow}
\usepackage{bm}
\usepackage{gensymb}  
\usepackage{xspace}   

\usepackage{svg}
\usepackage[colorlinks=true, urlcolor=blue, linkcolor=blue, citecolor=blue]{hyperref}

\usepackage[compat=1.0.0]{tikz-feynman}
\usepackage[caption=false]{subfig}

\usepackage{ulem} 

\usepackage{lineno}
\usepackage{amsmath}
\usepackage{etoolbox} 

\newcommand*\linenomathpatch[1]{%
  \cspreto{#1}{\linenomath}%
  \cspreto{#1*}{\linenomath}%
  \csappto{end#1}{\endlinenomath}%
  \csappto{end#1*}{\endlinenomath}%
}

\linenomathpatch{equation}
\linenomathpatch{gather}
\linenomathpatch{multline}
\linenomathpatch{align}
\linenomathpatch{alignat}
\linenomathpatch{flalign}

\begin{document}

\preprint{APS/123-QED}

\title{New Measurements of the Beam-Normal Single Spin Asymmetry in Elastic Electron Scattering Over a Range of Spin-0 Nuclei}

\collaboration{The PREX and CREX Collaborations}

\author{D.~Adhikari}\affiliation{Idaho State University, Pocatello, Idaho 83209, USA}
\author{H.~Albataineh}\affiliation{Texas  A  \&  M  University - Kingsville,  Kingsville,  Texas  78363,  USA}
\author{D.~Androic}\affiliation{University of Zagreb, Faculty of Science, Zagreb, HR 10002, Croatia}
\author{K.~Aniol}\affiliation{California State University, Los Angeles, Los Angeles, California  90032, USA}
\author{D.S.~Armstrong}\affiliation{William \& Mary, Williamsburg, Virginia 23185, USA}
\author{T.~Averett}\affiliation{William \& Mary, Williamsburg, Virginia 23185, USA}
\author{\mbox{C. Ayerbe Gayoso}}\affiliation{William \& Mary, Williamsburg, Virginia 23185, USA}
\author{S.~Barcus}\affiliation{Thomas Jefferson National Accelerator Facility, Newport News, Virginia 23606, USA} 
\author{V.~Bellini}\affiliation{Istituto  Nazionale  di  Fisica  Nucleare,  Sezione  di  Catania,  95123  Catania,  Italy}
\author{R.S.~Beminiwattha}\affiliation{Louisiana Tech University, Ruston, Louisiana 71272, USA}
% ...

\date{\today}

\begin{abstract}
We report precision determinations of the beam normal single spin asymmetries ($A_n$) in the elastic scattering of 0.95 and 2.18~GeV electrons off $^{12}$C, $^{40}$Ca, $^{48}$Ca, and $^{208}$Pb at very forward angles where the most detailed theoretical calculations have been performed. The first measurements of $A_n$ for $^{40}$Ca and $^{48}$Ca are found to be similar to that of $^{12}$C, consistent with expectations thus demonstrating the validity of theoretical calculations for nuclei with Z~$\leq20$.
We also report $A_n$ for $^{208}$Pb at two new momentum transfers (Q$^2$) extending the previous measurement. Our new data confirm the surprising result previously reported, with all three data points showing significant disagreement with the results from the $Z\leq 20$ nuclei.
These data confirm our basic understanding of the underlying dynamics that govern $A_n$ for nuclei containing $\lesssim 50$ nucleons, but point to the need for further investigation to understand the unusual $A_n$ behaviour discovered for scattering off $^{208}$Pb.
 
\end{abstract}

\maketitle

\newcommand{\he}{$^{4}\textrm{He}$\xspace}
\newcommand{\C}{$^{12}\textrm{C}$\xspace}
\newcommand{\pb}{$^{208}\textrm{Pb}$\xspace}
\newcommand{\al}{$^{27}\textrm{Al}$\xspace}
\newcommand{\si}{$^{28}\textrm{Si}$\xspace}
\newcommand{\zr}{$^{90}\textrm{Zr}$\xspace}

In the scattering of polarized electrons off nuclei, a well-known relativistic effect arises when there is any significant transverse polarization i.e. a polarization component in the plane  perpendicular to the incident electron momentum. In the electron's rest frame, the scattering amplitude then develops an azimuthal dependence due to the interaction of the electron's magnetic moment with the moving target nucleus's magnetic field, the latter being proportional to the electron's orbital angular momentum. 

The experimental observable that quantifies the size of the azimuthal modulation is the beam-normal single spin asymmetry (BNSSA), previously also called the vector analyzing power. This quantity is defined as the fractional difference in the scattering cross section when the electron polarization direction is reversed after being set up to be 100\%\ transverse. 
The BNSSA ($A_n$) depends on the scattering angle ($\theta$) and beam energy ($E_{\rm beam}$). The observed asymmetry is 
\begin{equation}
    A(\phi) = A_n (\theta,E_{\rm beam}) P_n \cos(\phi),
\end{equation}
where $\phi$ is the angle between the incoming beam polarization vector and the normal vector to the scattering plane. If $\vec{P}$ is the electron polarization and $\hat{n}$ is the unit vector normal to the scattering plane given by ${(\vec{k}\times\vec{k'})}/(|\vec{k}\times\vec{k'}|)$ with $\vec{k} (\vec{k'})$ the momentum of the incoming (outgoing) electron, then $P_n \cos(\phi) =  \vec{P}\cdot\hat{n} $.

For ultra-relativistic electron scattering, the dynamics of $A_n$ can involve internal structure of the target nucleus. 
Since time-reversal invariance dictates that $A_n$ should vanish at first order Born approximation, a non-zero measurement of $A_n$ is a probe of higher order effects (such as the exchange of multiple virtual photons).
The dominant contribution is the interference between the imaginary part of the two-photon exchange (TPE) amplitude and the one-photon exchange amplitude~\cite{DeRujula:1972te}. 

In recent years, theoretical calculations of electron scattering amplitudes have had to incorporate the exchange of one or more additional photons in order to interpret precision data on elastic form factors~\cite{Carlson:2007sp}. Such improvements are technically challenging but theoretically  interesting since one must evaluate contributions from the full range of off-shell intermediate states of the target nucleus. Indeed, $A_n$ measurements are an  important testing ground for theoretical calculations since the TPE amplitude is the dominant contribution for this observable.

Experimental measurements of $A_n$ have become feasible and have taken on additional significance because of its importance in the interpretation of weak neutral current (WNC) interaction measurements. Since the late 1970's, the technique of ultra-relativistic, longitudinally polarized electrons scattering off fixed targets has been used to measure the WNC interaction  between electrons and target nuclei, or their constituents, mediated by the Z$^0$ boson~\cite{Maas:2017snj,Kumar:2013yoa,Erler:2014fqa}. 
The observable is the parity-violating asymmetry $A_\text{PV}$, defined as the fractional difference in the scattering cross section for incident right(left)-handed electrons. Since $A_\text{PV}$ is predominantly sensitive to the ratio of the WNC amplitude to the electromagnetic amplitude, its size ranges from $10^{-7}$ to $10^{-4}$. 

In $A_\text{PV}$ measurements, $A_n$ leads to a spurious asymmetry when the electron beam polarization is not perfectly longitudinal. The size of the dominant contribution in terms of the fine-structure constant $\alpha_\mathrm{em}$, the electron mass $m_e$ and incident electron beam energy $E_e$ is $A_n\sim \alpha_\mathrm{em} m_e/E_e$. Under the conditions of typical $A_\text{PV}$ experiments $A_n$ ranges from $10^{-6}$ to $10^{-4}$, often much bigger than $A_\text{PV}$. The program of $A_\text{PV}$ measurements necessarily includes auxiliary $A_n$ measurements to ensure accurate corrections. A significant amount of $A_n$ data over a range of nuclei have thus become available.

One approach to $A_n$ calculations for electron-nucleon scattering has been to employ the optical theorem to relate the virtual photoabsorption cross section to the doubly-virtual Compton scattering amplitude~\cite{Afanasev:2004pu,Gorchtein:2004ac,Gorchtein:2005yz,Gorchtein:2005za}.  This approach intrinsically includes all excited intermediate states but is only valid in the forward limit. 
Another approach takes into account intermediate states via a parametrization of electroabsorption amplitudes~\cite{Pasquini:2004pv,Tomalak:2016vbf}.  In this case, the excited hadronic states are limited to the $\pi N$ states that have been experimentally measured, but the calculation is valid at all angles.  
Finally, heavy baryon chiral perturbation theory, though valid only at low incident energy, has been used~\cite{Diaconescu:2004aa}.

The measurements of $A_n$ on the proton~\cite{Wells:2000rx,Armstrong:2007vm,Androic:2011rh,Maas:2004pd,Gou:2020viq,Rios:2017vsw,Androic:2020rkw} span beam energies from 0.3 to 3 GeV and a large range of lab scattering angles from far forward to backward.  The larger angle data generally do not agree well with calculations, demonstrating the importance of including intermediate states beyond the $\pi N$ states.  However, at forward angles, the data are in better agreement with calculations based on the optical theorem approach.

The HAPPEX and PREX collaborations expanded the study of $A_n$ to nuclei, with measurements on \he, \C, and \pb at forward angle, 6\degree, and energies 1~-~3~GeV~\cite{Abrahamyan:2012cg}.  
The A1 collaboration reported results on \C with a beam energy of 570 MeV and larger angles (15\degree--26\degree) for a variety of $Q^2$~\cite{Esser:2018vdp}, and from $^{28}$Si and $^{90}$Zr with the same energy and similar angles~\cite{Esser:2020vjb}.  The Q$_{\rm weak}$ collaboration reported results on \C and \al at an angle of 7.7$^\circ$~\cite{PhysRevC.104.014606}.

Two different approaches have been undertaken to extend $A_n$ calculations beyond scattering off nucleons. The first approach numerically solves the Dirac equation for the electron moving in the Coulomb field of an infinitely heavy nucleus in the distorted-wave approximation but neglects excited intermediate states~\cite{Cooper:2005sk}.  
These calculations significantly under-predict the data.  The second approach extends the optical theorem calculations to nuclei~\cite{Afanasev:2004pu,Gorchtein:2008dy}. In Ref.~\cite{Gorchtein:2008dy} this is done by approximating the  photoabsorption cross section for nuclei as the photoabsorption cross section for a single nucleon, scaled by the mass number $A$.  The $Q^2$ dependence is included using the known charge form factor for each nucleus and a single Compton slope parameter for all nuclei. These optical model calculations approximately describe all the existing (forward) nuclear data within uncertainties~\cite{Esser:2018vdp,Esser:2020vjb,PhysRevC.104.014606}, with the exception of $^{208}$Pb.

The precision $A_n$ measurements reported by the HAPPEX and PREX collaborations~\cite{Abrahamyan:2012cg} were of particular interest since all the conditions for a robust comparison to theory were satisfied, namely, high incident beam energy, very forward scattering angle, and clean separation of inelastically scattered electrons. These measurements on $^{1}$H, $^{4}$He, and $^{12}$C agree with theoretical predictions, whereas the measurement on $^{208}$Pb showed a substantial disagreement (so-called ``PREX puzzle''); indeed $A_n$ was consistent with zero within quoted uncertainties. 

A popular speculation for the PREX puzzle has been the inadequacy of theoretical calculations to simultaneously account for Coulomb distortions and disperson corrections in a heavy nucleus. Recently, Koshchii~{\it et al.} have performed more sophisticated calculation of $A_n$ for heavy nuclei using a hybrid of the two previous approaches~\cite{Koshchii:2021mqq}.  The Dirac equation is solved numerically and the contribution of inelastic intermediate states is included in the form of an optical potential with an absorptive component. In addition, the Compton slope parameter is made $A$-dependent based on experimental data.  

In this Letter we report new measurements of nuclear $A_n$ with similar kinematic conditions to the previous HAPPEX/PREX measurements, including the first measurements from $^{40}$Ca and $^{48}$Ca targets. 
The data were obtained in auxiliary measurements during the PREX-2~\cite{Adhikari:2021phr} and CREX~\cite{Horowitz:2013wha} experiments, which ran consecutively in 2019 and 2020, and used the same experimental apparatus with 0.95~GeV and 2.18~GeV electron beam energy respectively.  The primary goal of these experiments was to measure $A_\text{PV}$ for $^{208}$Pb and $^{48}$Ca using a longitudinally polarized electron beam.  
Special data taking with transversely polarized beam yielded BNSSA measurements for the primary targets $^{208}$Pb and $^{48}$Ca, as well as $^{12}$C and $^{40}$Ca. Experimental techniques, refined for greater precision and cross-checked with larger data sets in the $A_\text{PV}$ measurements, were more than sufficient to provide the required precision and control of possible systematic uncertainties for the BNSSA studies. 

Longitudinally polarized electrons were produced in the CEBAF injector using circularly polarized laser light incident on a strained GaAs photocathode. 
The extracted electron helicity was rapidly flipped by reversing the sign of the laser's circular polarization with a Pockels~\cite{Sinclair:2007ez} cell at a frequency of 120 or 240 Hz. The BNSSA measurements required a special configuration in which the polarization of the electron beam was rotated to vertical using a Wien filter and solenoids~\cite{Sinclair:2007ez}. 
A half-wave $\lambda/2$ plate was used periodically to reverse the laser polarization independently from the Pockels cell control. By accumulating roughly equal statistics in the two slow-reversal states many systematic effects were thereby suppressed.

The isotopically enriched $^{208}$Pb, $^{48}$Ca, $^{40}$Ca, and $^{12}$C targets (thicknesses of 625, 990, 1000 and 458~mg/cm$^2$, respectively) were mounted in a copper frame cooled with 15K He gas. For thermal stability and efficient cooling, the $^{208}$Pb was sandwiched  between two 90~mg/cm$^2$ diamond foils and the beam was rastered over each target.
Electrons elastically scattered from the target, with symmetric acceptance on either side of the beam, were focused onto the focal plane of the two Hall A High Resolution Spectrometers (left- and right- HRS)~\cite{Alcorn:2004sb}, where they were intercepted by a pair of  $16\times3.5\times0.5$~cm$^3$ fused silica (quartz) detectors. 
The HRS momentum resolution ensured that nearly all accepted events were due to elastic scattering.

A relative measure of the scattered flux was obtained by collecting the quartz Cherenkov light in a photomutiplier tube (PMT) whose output was integrated over a fixed time period of constant helicity (``helicity windows'') and digitized. 
The raw asymmetry $A_\text{raw}$ is the fractional difference in the integrated detector response normalized to the beam intensity for two opposite helicity states. 
Window octets (for 240~Hz helicity flip) with the pattern $+--+-++-$ or its complement (choice made by a pseudorandom number generator) were used to suppress 
60~Hz power line noise. 

\begin{figure}[ht]
    \centering
    \includegraphics[width=0.45\textwidth]{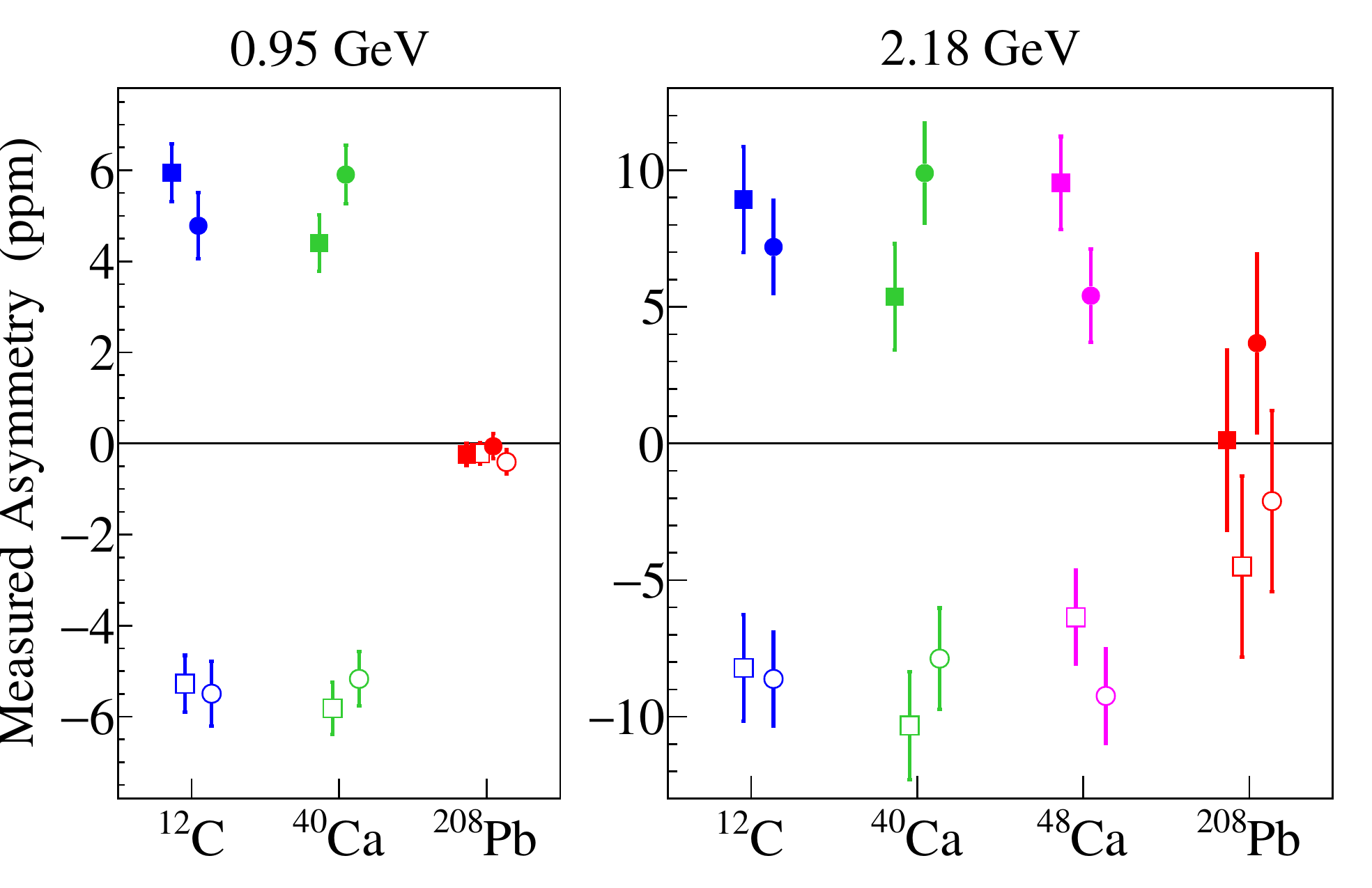}
    \caption{Measured asymmetries, corrected for beam fluctuations and sign-corrected for slow helicity reversals, demonstrating consistency over 4 configurations.  Data from the left (right) HRS is shown with filled (open) symbols, while circles (squares) represent the half-wave $\lambda/2$ plate in (out) configuration.}
   \label{fig:Andata}
\end{figure}

Noise contributions from fluctuations in the beam 
characteristics in each helicity window tend to be substantial due to the very forward kinematics and steeply falling nuclear form factor as a function of scattering angle. The beam correction is defined as $A_\text{beam}=\sum{c_i \Delta X_i}$, where $\Delta X_i$ is the measured parameter difference and $c_i$ are the detector sensitivities to fluctuations in beam energy, position and angle. The detector sensitivities are determined in two independent ways. The first uses beam modulations in which the beam parameters (position, angle, and energy) are varied independently over the phase-space of possible beam jitter. The system was operated several times per hour using air-core steering coils and an accelerating cavity to modulate the energy. Beam monitors in Hall A provided measurements of the beam parameters during these modulation periods as well as during normal running.  The second uses a linear regression of sensitivity slopes measured during normal running, which were consistent with those determined with the modulation data.

The measured asymmetry  $A_\text{meas}=A_\text{raw}-A_\text{beam}$ averages for all targets are shown separately
for each spectrometer in Fig.~\ref{fig:Andata}, after sign correcting for the half-wave ($\lambda/2$) plate reversals. The asymmetries are statistically similar in magnitude but opposite in sign for scattering to 
the left or the right, as expected. We therefore average the asymmetry difference between the left- and right-arm detectors (the double-difference asymmetry) for window pairs in the final analysis. 

We note that the dominant source of beam noise arose from position fluctuations in the horizontal direction, which change the acceptance of the spectrometers in opposite directions.
Noise in the beam energy or current largely cancels in the double-difference asymmetry. 
The corrections, typically comparable in magnitude to the respective statistical uncertainties in $A_\text{meas}$,  were extracted from the beam modulation data. Based on the understanding of the correction techniques developed in the larger $A_\text{PV}$ data sets, a conservative estimate of 5\% uncertainty is assigned on individual corrections for each beam parameter. 

The corrected asymmetry, $A_\text{corr}$, takes into account the contributions from each background process, $i$, and is given by
\begin{equation}
  A_\text{corr}=\frac{A_\text{meas} - \sum{f_i A_i}}{1 - \sum{f_i}},  
\end{equation}
where $f_i$ is the rate fraction of the process, and $A_i$ is the asymmetry of that process.
The fractional rate ($R_\text{C}/(R_\text{Pb}+R_\text{C})$) from the carbon in the diamond backing of the $^{208}$Pb target was estimated via simulation and found to be 0.063~$\pm$~0.006 at 0.95~GeV and 0.38~$\pm$~0.06 at 2.18~GeV.
The $^{40}$Ca contribution to the $^{48}$Ca measurement (at 2.18~GeV) was estimated to be $R_{^{40}\text{Ca}}/(R_{^{48}\text{Ca}}+R_{^{40}\text{Ca}}) =0.0831$~$\pm$~0.0017. 
Other impurities in the $^{48}$Ca target were below 0.05\% atomic percent and were neglected.

The kinematic distributions of the HRS acceptance were obtained in dedicated low  beam-current measurements where individual particles were tracked using Vertical Drift Chambers~\cite{Alcorn:2004sb} located near the focal plane. The average four-momentum transfer $\langle Q^2 \rangle$ was determined to an accuracy of $<1\%$. 
The $\langle Q^2 \rangle$ differences between left and right arm were approximately $0.5\%$ and $2.5\%$ for the PREX-2 and CREX configurations respectively. The average of the left and right $Q^2$ values for each target are shown in Table \ref{tab:Kinematics}. These data also yielded a small $\langle\cos{\phi}\rangle$ correction since the spectrometer acceptance is not confined to the horizontal plane; it varies slightly depending on the scattered flux distribution of each target.
While the experimental setup and HRS momentum resolution ensured that the acceptance from inelastic scattering was negligible for 0.95~GeV data, a more careful analysis was required for 2.18~GeV data, where the largest accepted contamination (1.3\%) came for the $^{40}$Ca $3^-$ state. The BNSSA associated with the inelastic events was assumed to be the same as for elastic events~\cite{PhysRevC.104.014606} and was assigned 100\% uncertainty.

\begin{table}[ht]
    \centering
    \begin{tabular}{c c c c c}
      \toprule 
         E$_{\rm beam}$ (GeV) & Target   & $\langle\theta_\text{lab}\rangle$~(deg)  & $\langle Q^2 \rangle$~(GeV$^2$)  &  $\langle\cos{\phi}\rangle$ \\
     \hline
           0.95 & $^{12}$C   & 4.87  & 0.0066  & 0.967 \\
           0.95 & $^{40}$Ca  & 4.81  & 0.0065  & 0.964 \\
           0.95 & $^{208}$Pb & 4.69  & 0.0062  & 0.966 \\ \\
           
           2.18 & $^{12}$C   & 4.77  & 0.033  & 0.969 \\
           2.18 & $^{40}$Ca  & 4.55  & 0.030  & 0.970 \\
           2.18 & $^{48}$Ca  & 4.53  & 0.030  & 0.970 \\
           2.18 & $^{208}$Pb & 4.60  & 0.031  & 0.969 \\
     \bottomrule      
    \end{tabular}
    \caption{$A_n$ measurement kinematics.}
    \label{tab:Kinematics} 
\end{table}

For the 0.95~GeV data, the transverse beam polarization was determined to be $P_n = 89.7 \pm 0.8\%$ using dedicated M\o ller polarimeter longitudinal polarization measurements.  For the 2.18~GeV data, both Compton and M\o ller polarimeters were operational and yielded consistent results; the combined result obtained was $P_n = 86.8 \pm 0.7\%$. Using these beam polarizations we finally obtain $A_n$ as: 
\begin{equation}
  A_{\rm n} = \frac{A_\text{corr}}{P_n \langle\cos{\phi}\rangle}.  
\end{equation}

\begin{table}[ht]
    \centering
    \begin{tabular}{cccc|cccc}
      \toprule
        E$_\text{beam}$ & \multicolumn{3}{c}{0.95~GeV}  &   \multicolumn{4}{c}{2.18~GeV} \\ 
        Target & $^{12}$C & $^{40}$Ca & $^{208}$Pb & $^{12}$C & $^{40}$Ca & $^{48}$Ca & $^{208}$Pb\\
     \midrule
        $A_\text{false}$ & 0.03 & 0.05 & 0.08 & 0.04 &0.06 & 0.10 & 0.03\\
        Polarization & 0.06 & 0.05 & $<$0.01 & 0.08 &0.08 & 0.08 & $<$0.01\\
        Non-linearity & 0.03 & 0.03 & $<$0.01 & 0.05 &0.05 & 0.05 & 0.01\\
        Tgt. impurities & $<$0.01 & $<$0.01 & 0.04 & $<$0.01 &$<$0.01 & 0.10 & 0.80\\ 
        Inelastic & $<$0.01 & $<$0.01 & $<$0.01 & 0.08 &0.15 & 0.08 & $<$0.01\\ \\
     
           Tot. Syst. & 0.07 & 0.08 & 0.09 & 0.13 &0.18 & 0.19 & 0.75\\
          Statistical & 0.38 & 0.34 & 0.16 & 1.05 &1.10 & 1.09 & 3.15\\ \\
     
          Tot. Unc. & 0.39 & 0.34 & 0.18 & 1.05 &1.11 & 1.11 & 3.23\\
          \bottomrule
    \end{tabular}
    \caption{$A_n$ measurement uncertainty contributions in units of $10^{-6}$ (ppm).}
    \label{tab:Uncertainties} 
\end{table}

 A contribution from non-linear response of the PMT for each quartz detector was bounded to be $<0.5\%$ in bench tests. A summary of the main contributions to the uncertainties for the various targets is shown in Table~\ref{tab:Uncertainties}. The statistical uncertainties typically dominate and the systematic uncertainties are well under 1 ppm.

\begin{table}[ht]
    \centering
    \begin{tabular}{ccccc}
      \toprule 
          E$_{\rm beam}$ &  \multirow{2}{*}{Target} & \multirow{2}{*}{$\rm A_n$ (ppm)} & \multirow{2}{*}{$A_\text{avg}^{Z\leq20}$ (ppm)} & \multirow{2}{*}{$\frac{\rm A_n - A_{avg}^{Z\leq20}}{\rm uncert}$}  \\
           ${\rm (GeV)}$ & & & & \\
     \midrule
          0.95 & $^{12}$C   & \multirow{2}{*}{$\left.\begin{array}{c}
          \ -6.3\pm0.4 \\
          \ -6.1\pm0.3 
          \end{array}\right\rbrace$}   &\multirow{2}{*}{$-6.2\pm0.2$} & \\
          0.95 & $^{40}$Ca  &  & & \\
     \midrule
          0.95 & $^{208}$Pb & $\ 0.4\pm0.2$  & & 21~$\sigma$ \\ \\ 

          2.18 & $^{12}$C   &  \multirow{2}{*}{$\left.\begin{array}{c}
          \ \ -9.7\pm1.1 \\
          \ -10.0\pm1.1 \\
          \ \ \ -9.4\pm1.1 
          \end{array}\right\rbrace$}    & \multirow{3}{*}{$-9.7\pm0.6$}& \\
          2.18 & $^{40}$Ca  &      & & \\
          2.18 & $^{48}$Ca  &      & & \\
         \midrule
          2.18 & $^{208}$Pb &  \ \ \ $0.6\pm3.2$ & & 3.2~$\sigma$ \\
     \bottomrule
    \end{tabular}
    \caption{$A_n$ results for the four nuclei along with the corresponding total uncertainties (statistical and systematic uncertainties combined in quadrature).}
    \label{tab:Results} 
\end{table}

\begin{figure}[ht]
    \centering
    \includegraphics[width=\linewidth]{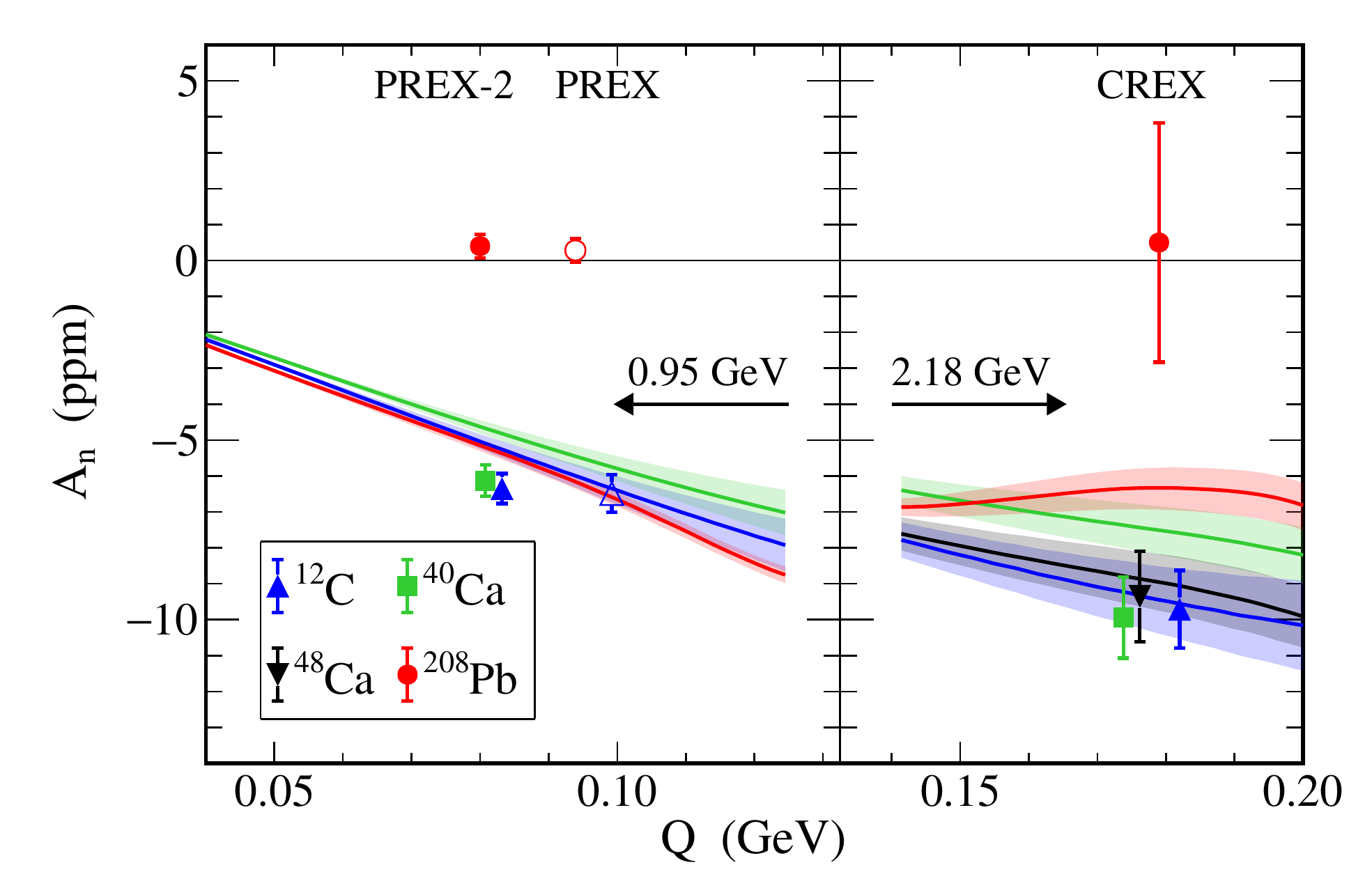}
    \caption{$A_n$ measurements from PREX-2, PREX (open circle and triangle, previously published~\cite{Abrahamyan:2012cg}) and CREX, at beam energies of 0.95\;GeV, 1.06\;GeV and 2.18\;GeV respectively. The solid lines show theoretical calculations from~\cite{Koshchii:2021mqq} at 0.95\;GeV and 2.18\;GeV together with their respective one sigma uncertainty bands. The color of each band represents the calculation for the same color data point. 
     Overlapping points are offset slightly in $Q$ to make them visible. 
    }
    \label{fig:Antheory}
\end{figure}

 Our final results are shown in Table~\ref{tab:Results}.
They are also displayed in Fig.~\ref{fig:Antheory}, with theoretical prediction~\cite{Koshchii:2021mqq} curves overlaid. It can be seen that the data are consistent with the previously published PREX results (open symbols) ~\cite{Abrahamyan:2012cg}.
Of particular note is that, at each beam energy, measurements on multiple nuclei with $Z\leq 20$ are consistent with each other within quoted uncertainties. 
Using a simple average on all but the $^{208}$Pb measurement we observe that the measured $A_n$ for $^{208}$Pb at 0.95~GeV is different by 21 standard deviations. Following a similar procedure for 2.18~GeV data we obtain 3.2 standard deviations.
Measurements to date support this simple averaging procedure provided that the three conditions mentioned earlier are satisfied, namely, high incident beam energy, very forward scattering angle, and clean separation of inelastic scattered electrons.
It is worth noting that the $^4$He data published by the HAPPEX collaboration~\cite{Abrahamyan:2012cg} taken at 2.75~GeV beam energy, scaled by $\sqrt{Q^2}$ to the same kinematics as our 2.18~GeV data, is consistent with the average presented in Table~\ref{tab:Results}.

The original PREX data challenged the community to explain an asymmetry for $^{208}$Pb which is an order of magnitude smaller than for other nuclei and theoretical expectations.  We have now shown that this effect occurs over a range of beam energies and $Q^2$, likely ruling out an explanation based on the location of a diffractive minimum. A recent calculation has, for the first time, included both effects of Coulomb distortion and excited intermediate states~\cite{Koshchii:2021mqq}. While the resulting prediction moved closer than previous calculations to the 2.18~ GeV $^{208}$Pb $A_n$ measurement, the disagreement remains stark at 0.95~GeV - a firm indication that further theoretical investigation is warranted. 

It is especially difficult to explain the small $A_n(^{208}Pb)$ given our new results showing good agreement for $Z = 20$ nuclei.  A theoretical correction is required that not only yields $A_n(^{208}Pb)\approx 0$ for a significant range of $Q$ while being small for $Z\leq 20$ nuclei but also goes beyond dynamics included in the calculations of ref.~\cite{Koshchii:2021mqq}. 

One such possible radiative correction is a vertex correction to the nucleus: a Feynman diagram where a virtual photon connects the initial and final state legs of the nucleus. Such a correction would naively be of the  order $Z^2 \alpha$ since it contains an additional photon coupling to a charge $Z$ nucleus, compared to the leading TPE diagram. For Pb this is very large, $Z^2\alpha=49.1$. This correction to $A_n$ would take the form,

$$A_\text{n} \approx A_\text{0}(\text{Q})(1 - \text{C} \cdot \text{Z}^2\alpha),$$ 
where $A_\text{0}(\text{Q})$ is a theoretical prediction for $A_\text{n}$ without the radiative correction or the experimentally measured asymmetry for low Z nuclei,  and $C$ is an empirical constant.  We find $C\approx 0.02$ in order to reproduce the small $A_n(^{208}$Pb$)$ values without degrading the agreement for $Z \leq 20$.
We find that the $^{90}$Zr $A_n$ central values~\cite{Esser:2020vjb} from Mainz are consistent with such a correction for the same value of $C$.
Higher precision data for $^{90}$Zr as well as for other suitable nuclei with $Z\geq 40$ would allow a test of this hypothesis.

In conclusion, we have reported new measurements of $A_n$ for $^{12}$C, $^{40}$Ca, $^{48}$Ca and $^{208}$Pb. 
The $A_n$ measurements on light and intermediate mass nuclei are consistent with each other and extend the range of nuclei for which a robust theoretical comparison can be made up to Z~$=20$. On the other hand, these measurements are in significant disagreement with the $^{208}$Pb measurements, confirming and extending the previously reported PREX result suggesting that the suppression of the asymmetry is $Q^2$ independent. The level of disagreement in $A_n$ is far beyond that expected from current theoretical models and merits further experimental and theoretical investigation.

\begin{acknowledgments}
We thank the entire staff of JLab for their efforts to develop and maintain the polarized beam and the experimental apparatus, and acknowledge the support of the U.S. Department of Energy, the National Science Foundation and NSERC (Canada). This material is based upon the work supported by the U.S. Department of Energy, Office of Science, Office of Nuclear Physics Contract No. DE-AC05-06OR23177. 
\end{acknowledgments}

\bibliography{Transverse}

\end{document}